\documentclass[article]{ajs}

\usepackage{amsmath}
\usepackage{amsfonts}
\usepackage{algorithm}
\usepackage{enumerate}
\usepackage{algpseudocode}
\usepackage{listings}
\usepackage{bm}

\newcommand{\Bgamma}{\boldsymbol{\mathsf{\gamma}}}
\newcommand{\By}{\boldsymbol{Y}}
\newcommand{\Bx}{\boldsymbol{X}}
\newcommand{\Bchi}{\boldsymbol{\chi}}

\author{Aliaksandr Hubin\\ Norwegian Computing\\ Center\\Dep.Mathematics\\ University of Oslo\And 
        Geir O. Storvik\\ Dep.Mathematics\\ University of Oslo\\ Norwegian Computing\\ Center\And
        Paul E. Grini\\Dep. Biosciences\\ University of Oslo\And
        Melinka A. Butenko\\Dep. Biosciences\\ University of Oslo}
\title{A Bayesian binomial regression model with latent Gaussian processes for modelling DNA methylation}

\Plainauthor{Aliaksandr Hubin,  Geir O. Storvik, Paul E. Grini,  Melinka A. Butenko} %% comma-separated
\Plaintitle{A Bayesian binomial regression model with latent Gaussian processes for modelling DNA methylation} %% without formatting
\Shorttitle{A Bayesian binomial regression model with latent GP for modelling DNA methylation} %% a short title (if necessary)

\Abstract{
Epigenetic observations are represented by the total number of reads from a given pool of cells and the number of methylated reads, making it reasonable to model this data by a binomial distribution. There are numerous factors that can influence the probability of success in a particular region. Moreover, there is a strong spatial (alongside the genome) dependence of these probabilities. We incorporate dependence on the covariates and the spatial dependence of the methylation probability for observations from a pool of cells by means of a binomial regression model with a latent Gaussian field and a logit link function. We apply a Bayesian approach including prior specifications on model configurations. We run a mode jumping Markov chain Monte Carlo algorithm (MJMCMC) across different choices of covariates in order to obtain the joint posterior distribution of parameters and models. This also allows finding the best set of covariates to model methylation probability within the genomic region of interest and individual marginal inclusion probabilities of the covariates. 
}
\Pages{1--xx}

\Address{
  Aliaksandr Hubin\\
  Department of Statistical Analysis, Machine Learning and Image Analysis\\
  Norwegian Computing Center\\
  P.O. Box 114 Blindern, NO-0314 Oslo, Norway\\
  E-mail: \email{aliaksandr.hubin@nr.no}\\
  URL: \url{https://www.nr.no/?q=publicationprofile&query=ahu/}
}
\begin{document}

%% include your article here, just as usual

\section{Introduction}

Epigenetic modifications contribute to the generation of phenotypic plasticity, but the understanding of its contribution to phenotypic alterations and how the genome influences epigenetic variants requires
%Natural epigenetic variation provides a source for the generation of phenotypic diversity, but
%to understand its contributions to such diversity and its interaction with genetic variation requires
further investigation \citep{schmitz2013patterns}. Epigenetic changes are crucial for the development and differentiation of various
cell types in an organism, as well as for normal cellular processes. 
%Epigenetic modifications act as regulatory mechanisms and if they occur in the promoter of the corresponding gene they suppress gene transcription. 
Epigenetic modifications modulate gene expression and modifications found in the promoter or regulatory elements play a prominent role in activating or suppressing transcript levels.
This creates interesting research possibilities, which are often challenging from the statistical point of view. For example, \citet{li2019bayesian} suggested a Bayesian negative binomial regression model to study the influence of methylation (used as covariates) on RNA-Seq gene expression counts (used as observations).  Also, \citet{tang2017integration} developed a Gaussian Bayesian regression model to link the differential gene expression (measured as log$_2$ fold change) to various exogenous variables including tumour suppressor genes categories, mean methylation values and genomic segment distributions. In turn, \citet{ma2017multiple} suggested a multiple network for epigenetic studies and implemented the Cox proportional hazard model to analyze the association of methylation profile of each epigenetic module with the patient survival. Recently, high-throughput epigenetics experiments have enabled researchers to measure genome-wide epigenetic profiles. This allows performing Epigenome-wide association studies (EWAS), which also hold promise for the detection of new regulatory mechanisms that may be susceptible to modification by environmental and lifestyle factors~\citep{michels2013recommendations}.

A major task today is the development of models and statistical methods for linking epigenetic patterns to genomic and/or environmental variables and interpreting them. Unlike the papers mentioned above \citep{ma2017multiple, tang2017integration, li2019bayesian}, we use methylation data as responses and link them to genomic and phenotypic variables (used as covariates). Moreover, by means of performing careful statistical modelling, our model takes into account that epigenetic data are spatially correlated (along locations in the genome) with high noise levels. Due to the availability of the data, our focus will be on the model plant \emph{Arabidopsis thaliana}. For instance, \citet{becker2011spontaneous} previously analysed Arabidopsis data consisting of epigenetic observations on a set of 10 lines, which were separately propagated in a common environment for 30 generations. These were compared with two independent lines propagated for only three generations. Their analysis aimed at global summaries of structures but was based on individual and (site-wise) hypothesis testing methods combined with false discovery rate control methodology.
% \citet{smith2013might} describe what is known about the roles of phenotypic plasticity and
% epigenetic variation in evolution and ecological speciation. 
% Phenotypic plasticity might be an important process during
% speciation as it may accelerate adaption to or persistence in a novel environment.
% Speciation is likely to start as a slow process with restricted gene flow at only a few key
% loci followed by a spread across the genome due to reduced gene flow influencing linked
% regions. Many studies indicate that epigenetic variation may prove to be a common mechanism underlying 
% environmentally induced phenotypes. Epigenetic mechanisms exists in almost every
% living organism and while virtually unstudied with respect to ecological speciation, these mechanishms have the potential to contribute to rapid adaptation to a changing environment. There is growing evidence that some epigenetic marks are heritable, suggesting that epigenetically controlled plasticity could carry over into 
% subsequent generations. \cite{schmitz2013patterns} made analysis on \emph{Arabidopsis} with focus on
% single methylation polymorphisms (SMPs) from genotypical distinct wild accessions in order to take genetic variants into account. 
In this paper, however, we limit ourselves to finding a pattern of signals appearing along the single genome that significantly influences the methylation probability of the corresponding organism. 
%We additionally take into account spatial dependence between the observations as well as the unexplained by the exogenous variables variability of the epigenetic observations. 
This is done by means of applying a binomial regression model with latent Gaussian variables, which take into account both spatial dependence and variability that can not be explained by the exogenous variables alone. The chosen latent Gaussian variable is a sum of a random walk $RW(1)$ component and an independent $IG$ component. Model selection and parameter estimation within models are performed simultaneously
in a Bayesian framework, applying the mode jumping Markov chain Monte Carlo (MJMCMC) algorithm developed by \citet{Hubin2016} to perform the computations involved.
%to the Bayesian binomial regressionwith a random walk of order one, denoted as $RW(1)$, and independent Gaussian, denoted as $IG$, latent processes (empirical reasoning of choosing these mixture of latent Gaussians will is also given). 
MJMCMC outputs posterior model probabilities allowing to find the best combination of explanatory genomic variables and compute marginal inclusion probabilities for the importance of individual variables. Our approach also allows to generate a model-averaged classification of the methylation status at different locations and make imputations for those locations that do not have enough observations, whilst the currently used approach is to simply ignore these locations.

\section{Mathematical model}

We model the number of methylated reads $Y_t \in \{1,...,n_t\}$ per position in the genome (nucleotide base position) to be binomially distributed with the number of trials equal to the number of reads,  $n_t \in  \mathbb{N}$, for position  $t \in \{t_1,...,t_T\}$ (where $T$ is the total number of genomic positions in the addressed genomic region)  and corresponding probability of success $p_t \in \mathbb{R}_{[0,1]}$. The probability $p_t$ is modeled via the logit link to the covariates $\bm X_t = \{X_{t1},...,X_{td}\}$. These covariates might be a position within a gene (e.g promoter or coding region), an indicator of the underlying genetic structure, or other types of features (our choice of the covariates is given in Section~3). A latent Gaussian $RW(1)$ process $\{\delta_t\} \in \mathbb{R}$ and  a latent independent Gaussian ($IG$) process $\{\zeta_t\}$ are included into the model in order to take into account spatial dependence of methylation probabilities along the genome and the variance which is not explained by the covariates. Other explored latent Gaussian variables were also tested prior to variable selection on a full model before the selection of this structure was done (see the detail in the Appendix A of the paper). This gives the following model formulation:
\begin{align}
  \Pr(Y_t = y|n_t,  p_t) = & \binom {n_t} {y} p_t^{y}(1-p_t)^{n_t-y},\label{themodel} \\
  %p_t = & \frac{e^{\beta_0 + \sum_{j=1}^{d} \gamma_j\beta_{j}X_{tj} + \delta_t+\zeta_t}}{1+e^{\beta_0 + \sum_{j=1}^{d} \gamma_j\beta_{j}X_{tj} + \delta_t+\zeta_t}},\\
  \text{logit}(p_t) = & \beta_0 + \sum_{j=1}^{d} \gamma_j\beta_{j}X_{tj} + \delta_t+\zeta_t,\\
  \delta_t =  & \delta_{t-1} + \epsilon_t,\quad \epsilon_{t} \stackrel{iid}{\sim}  N(0,\tau_{\epsilon}^{-1}),\\
  \zeta_{t} \stackrel{iid}{\sim} & N(0,\tau^{-1}_{\zeta}),
\end{align}
where $\beta_j \in \mathbb{R}, j \in \{0,...,d\}$ are regression coefficients of the model showing whether and in which way the corresponding covariate influences the probability of methylation on average, $\gamma_j \in \{0,1\}, i \in \{1,...,d\}$ are latent indicators, defining if covariate $j$ is included into the model ($\gamma_j = 1$) or not ($\gamma_j = 0$), $\{\epsilon_t\}$ are the error terms of $RW(1)$ process $\{\delta_t\}$, which are normally distributed with zero mean and precision ${\tau^{-1}_\epsilon}$. Finally, $\tau^{-1}_{\zeta}$ is the precision term of the $IG$ process $\{\zeta_t\}$.  We then put the following priors for the parameters of the model:
\begin{align}
\gamma_i \sim  \text{Bernoulli}(q),
\beta_i|\gamma_i \sim  \text{I}(\gamma_i = 1) N(0,\tau^{-1}_{\beta}), \tau_{\beta},\tau_{\epsilon} \text{ and } \tau_{\zeta} \sim  \text{Gamma}(1,5\cdot 10^{-5}).\label{gammaprior}
\end{align}
Here, $q  = 0.5$ is the prior Bernoulli probability of including a covariate into the model, and $\text{I}(\cdot)$ is  the indicator function.

We perform analysis for the model defined by
Equations~\eqref{themodel}-\eqref{gammaprior} by means of the MJMCMC algorithm~\citep{Hubin2016}. The algorithm is capable of efficiently moving in the defined model space by means of both accurately exploring the modes of the probability mass and switching between these modes using large jumps combined with local optimization and randomization. 

\section{Bayesian inference}

Let $\Bgamma = (\gamma_1,...\gamma_p)$, which uniquely defines a specific model. Assuming the constant term $\beta_0$ is always included, there are $L = 2^{d}$ different models to consider. Define $\boldsymbol\theta = \{\boldsymbol{\beta},\boldsymbol{\psi},\tau_{\beta},\tau_{\epsilon},\tau_{\zeta}\} \in \Theta$, which describes the parameters of the models. Also let $\By=(y_1,...,y_n)$ and $\Bx=(\bm X_1,...,\bm X_n)$. We want to make inference jointly on models and their parameters  $p(\Bgamma,\boldsymbol{\theta}|\By,\Bx)$.  We also want to find a set of the best models with respect to posterior marginal model probabilities $p(\Bgamma|\By,\Bx)$. Finally, we want to obtain the marginal inclusion probabilities $p(\Bgamma_j=1|\By,\Bx),j \in \{1,...,d\}$ for individual covariates. 

By Bayes formula, $p(\boldsymbol{\gamma},\boldsymbol{\theta}|\By,\Bx) = p(\boldsymbol{\theta}|\boldsymbol{\gamma},\By,\Bx)p(\boldsymbol{\gamma}|\By,\Bx)$. In Section~\ref{sec:inla} we describe how to compute $p(\boldsymbol{\theta}|\boldsymbol{\gamma},\By,\Bx)$ and $p(\By|\Bx,\Bgamma)$. For the time being, we assume that marginal likelihoods $p(\By|\Bx,\Bgamma)$
are available for a given $\Bgamma$. Then by Bayes formula:
\begin{equation}\label{PMP}
p(\Bgamma|\By,\Bx) =  \frac{{p(\By|\Bx,\Bgamma)p(\Bgamma)}}{\sum_{\Bgamma' \in\Omega}{p(\By|\Bx, \Bgamma')p(\Bgamma')}}.
\end{equation}
In order to calculate $p(\Bgamma|\By,\Bx)$ we have to iterate through the whole model space $\Omega$, which becomes computationally infeasible for large $d$. 
The ordinary MCMC estimate is based on a number of MCMC samples $\Bgamma^{(i)},i= 1,...,W$:
\begin{equation}\label{map2}
\widetilde{p}(\Bgamma|\By,\Bx)=W^{-1}\sum_{i=1}^{W}{\text{I}(\Bgamma^{(i)} = \Bgamma)} \xrightarrow[W\rightarrow\infty]{d} p(\Bgamma|\By,\Bx).
\end{equation}
An alternative,  named the renormalized model (RM) estimates by \citet{Clyde:Ghosh:Littman:2010}, is
\begin{equation}
\widehat{p}(\Bgamma|\By,\Bx) =  \frac{{p(\By|\Bx,\Bgamma)p(\Bgamma)}}{\sum_{\Bgamma' \in \mathbb{V}}{p(\By| \Bx,\Bgamma')p(\Bgamma')}}\text{I}(\Bgamma \in \mathbb{V})\xrightarrow[\mathbb{V}\rightarrow\Omega]{d} p(\Bgamma|\By,\Bx),\label{approxpost}
\end{equation}
where now $\mathbb{V}$ is the set of visited models during the
MCMC (or any other model space exploration algorithm) run. Although both~\eqref{approxpost} and~\eqref{map2} 
are asymptotically consistent,  \eqref{approxpost} 
will often be a preferable estimator since convergence of the MCMC based approximation~\eqref{map2} 
is much slower, see~\citet{Clyde:Ghosh:Littman:2010, Hubin2016}.

We aim at approximating $p(\Bgamma|\By,\Bx)$ by means of searching for some subspace $\mathbb{V}$ of $\Omega$ making the approximation~\eqref{approxpost} as precise as possible.
Models with high values of $p(\By|\Bx,\Bgamma)p(\Bgamma)$ and regions of relatively high posterior mass are important to be included into $\mathbb{V}$. Missing them in $\mathbb{V}$ can introduce significant biases in our estimates. Note that these aspects are just as important for the standard MCMC estimate~\eqref{map2}. The difference is that while  the number of times a specific model is visited is important when using~\eqref{map2}, for~\eqref{approxpost} it is enough that the model is visited at least once. In this context, the denominator of \eqref{approxpost}, which we would like to be as high as possible, becomes an extremely relevant measure for the quality of the search in terms of being able to capture whether the algorithm visits all of the modes, whilst the size of $\mathbb{V}$ should be low in order to save computational time.

The marginal inclusion probability $p(\gamma_j=1|\By,\Bx)$ is defined as:
\begin{equation}
{p}(\gamma_j=1|\By,\Bx) = \sum_{\Bgamma' \in {\Omega}}{\text{I}(\gamma_j'=1)p(\Bgamma'|\By,\Bx).}
\end{equation}
It can be approximated either by the MCMC estimator:
\begin{equation}
    \widetilde{p}(\gamma_j=1|\By,\Bx)=W^{-1}\sum_{i=1}^{W}{\text{I}(\Bgamma^{(i)}_j = 1)} \xrightarrow[W\rightarrow\infty]{d} p(\gamma_j=1|\By,\Bx),
\end{equation}
or using the renormalized approach:
\begin{equation}\label{margininuspost}
\widehat{p}(\gamma_j=1|\By,\Bx) = \sum_{\Bgamma' \in \mathbb{V}}{\text{I}(\gamma_j'=1)p(\Bgamma'|\By,\Bx)}\xrightarrow[\mathbb{V}\rightarrow\Omega]{d} p(\gamma_j=1|\By,\Bx),
\end{equation}
giving a measure of importance for the covariates of the model.

\subsection{Integrated nested Laplace approximations}\label{sec:inla}

Within hierarchical models with latent Gaussian structures, integrated nested Laplace approximations (INLA) for efficient inference on the posterior distribution \citep{rue2009eINLA} can be used. Following the INLA terminology, we define $\bm z$ to be the set of latent Gaussian variables and the regression parameters $\bm\beta$ while $\bm\eta$ contains the remaining parameters (a low-dimensional vector).
%Consider the following partition of the parameters: $\boldsymbol\theta =(\boldsymbol\eta,\boldsymbol z)$, where $\boldsymbol z$ is  the vector of all the latent Gaussian variables and $\boldsymbol\eta$ is the vector of hyperparameters of $\boldsymbol z$. Here, the vector of slope coefficients $\boldsymbol\beta$ is a part of $\boldsymbol z$ allowing to make $\boldsymbol \eta$ low-dimensional, which is important to facilitate computations. 
The INLA approach is based on two steps. First the marginal posterior of the hyperparameters is approximated by
\begin{align}\label{eq:LaplaceHyper}
 p(\boldsymbol\eta|\By,\Bx,\Bgamma) \propto \frac{p({\boldsymbol z},\boldsymbol\eta,\By,\Bx,\Bgamma)}{ p(\boldsymbol z|\boldsymbol\eta,\By,\Bx,\Bgamma)} = 
\frac{p({\boldsymbol z},\boldsymbol\eta,\By,\Bx,\Bgamma)}{ \tilde{p}_G(\boldsymbol z|\boldsymbol\eta,\By,\Bx,\Bgamma)} \Big{|}_{{\boldsymbol z} = {\boldsymbol z}^*(\boldsymbol\eta)} + \mathcal{O}(T^{-{3}/{2}})\,.
\end{align}
Here,  $\tilde{p}_G(\boldsymbol z|\boldsymbol\eta,\By,\Bx,\Bgamma)$ is the Gaussian approximation of $p(\boldsymbol z|\boldsymbol\eta,\By,\Bx,\Bgamma)$, and $\boldsymbol z^*(\boldsymbol\eta)$ is the mode of the distribution $p({\boldsymbol z}|\boldsymbol\eta,\By,\Bx,\Bgamma)$. The posterior mode of the hyperparameters is found by maximizing the corresponding Laplace approximation using some gradient descent method (like for example the Newton-Raphson routine). Then an area with a relatively high posterior density of the hyperparameters is explored with either some grid-based procedure or variational Bayes. 

The second step involves the approximation of the latent variables for every set of the explored hyperparameters. Here, the computational complexity of the approximation depends on the likelihood type for the data $\By|\Bx$. If it is Gaussian, the posterior of the latent variables is Gaussian, and the approximation is exact and fully tractable. In the case the likelihood is skewed or heavy tails are present, a Gaussian approximation of the latent variables tends to become inaccurate and another Laplace approximation should be used:
\begin{align}\label{eq:laplaceLatent}
\tilde p_{\text{LA}}(z_i|\boldsymbol\eta, {\By,\Bx,\Bgamma})  \propto \frac{p({\boldsymbol z},\boldsymbol\eta, {\By,\Bx,\Bgamma})}{\tilde p_{\text{GG}}({\boldsymbol z}_{-i}|z_i,\boldsymbol\eta,{\By,\Bx,\Bgamma})}\Big{|}_{{\boldsymbol z}_{-i} = {\boldsymbol z}^*_{-i}(z_i,\boldsymbol\eta)}.
\end{align}
Here, $\tilde p_{\text{GG}}$ is the Gaussian approximation to $p({\boldsymbol z}_{-i}|z_i,\boldsymbol\eta,{\By,\Bx,\Bgamma})$ and ${\boldsymbol z}^*_{-i}(z_i,\boldsymbol\eta)$ is the corresponding posterior mode. The error rate of~\eqref{eq:laplaceLatent} is $\mathcal{O}(T^{-{3}/{2}})$. The full Laplace approximation of the latent fields defined in equation \eqref{eq:laplaceLatent} is rather time-consuming, hence more crude lower-order Laplace approximations are often used instead \citep[typically increasing the error rate to $\mathcal{O}(T^{-1})$,]                               []{tierney1986accurate}. Once the posterior distribution of the latent variables given the hyperparameters is approximated, the uncertainty in the hyperparameters can be marginalized out using the law of total probability \citep{rue2009eINLA}:
\begin{align}
\tilde p(z_i|{\By,\Bx,\Bgamma} ) = \sum_k \tilde p_{\text{LA}}(z_i|\boldsymbol\eta_k, {\By,\Bx,\Bgamma}) \tilde p(\boldsymbol\eta_k|{\By,\Bx,\Bgamma} ) \Delta_k,
\end{align}
where $\Delta_k$ is the area weight corresponding to the grid exploration of the posterior distribution of the hyperparameters.

\subsubsection{Computing the marginal likelihood}

The marginal likelihood is defined as follows: For data $\{\By,\Bx\}$ and model  $\Bgamma$, which includes some unknown parameters $\bm\theta$, the marginal likelihood is given by
\begin{equation}
p(\By|\Bx,\Bgamma)=\int_{\Theta}p(\By|\Bx,\bm\theta,\Bgamma)p(\theta|\Bgamma)d\bm\theta\label{mlikdef}
\end{equation}
where $p(\bm\theta|\Bgamma)$ is the prior for $\bm\theta$  under model $\Bgamma$ while $p(\By|\Bx,\Bgamma,\bm\theta)$ is the likelihood function conditional on $\bm\theta$. Again, consider $\boldsymbol\theta =(\boldsymbol\eta,\boldsymbol z)$.

INLA approximates marginal likelihoods by
\begin{equation}
\widetilde{p}(\By|\Bx,\Bgamma) = \int_{\boldsymbol Z} \left. \frac{p(\By,\boldsymbol z,\boldsymbol{\eta}|\Bx,\Bgamma)}{\tilde{\pi}_G(\boldsymbol{\eta}|\By,\Bx,\boldsymbol z,\Bgamma)} \right \rvert_{\boldsymbol{\eta} = \boldsymbol{\eta}^*(\boldsymbol z|\Bgamma)} d\boldsymbol z,
\end{equation}
where $\boldsymbol{\eta}^*(\boldsymbol z|\Bgamma)$ is some chosen value of $\boldsymbol{\eta}$, typically the posterior mode, while $\tilde{\pi}_G(\boldsymbol{\eta}|\By,\Bx,\boldsymbol z,\Bgamma)$ is a Gaussian approximation to $\pi(\boldsymbol{\eta}|\By,\Bx,\boldsymbol z,\Bgamma)$. The integration of $\bm z$ over the support ${\boldsymbol Z}$ can be performed by an empirical Bayes (EB) approximation or using numerical integration based on a central composite design (CCD) or a grid~\citep[see][for details]{rue2009eINLA}. 

\paragraph{A toy example on computing the marginal likelihood.} Consider an example from~\citet{NealBlog}, in which we assume the following model $\Bgamma$:
\begin{equation}
\begin{split}
Y|z,\Bgamma\sim& N(z,\tau_1^{-1});\quad
z|\Bgamma\sim N(0,\tau_0^{-1}).
\end{split}
\end{equation}
Then obviously the marginal likelihood is available analytically as
\begin{eqnarray}
&Y|\Bgamma \sim N(0,\tau_0^{-1}+\tau_1^{-1}),
\end{eqnarray}
and we have a benchmark to compare approximations to. The harmonic mean estimator~\citep{raftery2006estimating} is given by
\begin{equation*}
\tilde{p}(Y|\Bgamma) = \frac{W}{\sum_{i=1}^{W}\frac{1}{p(Y|z_i,\Bgamma)}}
\end{equation*}
where $z_i\sim p(z|Y,\Bgamma)$. This estimator is consistent, however, often requires unreasonably many iterations to converge. We performed the experiments with $\tau_1 = 1$ and $\tau_0$ being either 0.001, 0.1 or 10. The harmonic mean is obtained based on $W = 10^7$ simulations. 5 runs of the harmonic mean procedure are performed for each scenario. For INLA we used the default tuning parameters from the package \citep{rue2009eINLA}.
\begin{table}[ht]
\begin{center}
\begin{tabular}{c|c|c|c|c|rrrrr}
\hline
$\tau_0$&$\tau_1$&$T$&Exact&INLA&\multicolumn{5}{c}{H.mean}\\\hline
0.001&1&2&-7.8267&-7.8267&-2.4442&-2.4302& -2.5365&-2.4154&-2.4365\\\hline
0.1&1&2&-3.2463&-3.2463&-2.3130&-2.3248&-2.5177&-2.4193&-2.3960 \\\hline 
10&1&2&-2.9041&-2.9041&-2.9041&-2.9041&-2.9042& -2.9041&-2.9042\\\hline
\end{tabular}
\end{center}
\caption{Comparison of INLA, harmonic mean and exact marginal likelihood}\label{mlikcomp1}
\end{table}
As one can see from Table \ref{mlikcomp1}, INLA gives extremely precise results even for a huge variance of the latent variable, whilst the harmonic mean can often become extremely crude even for $10^7$ iterations. More examples showing the accuracy of INLA are summarized in~\citet{HubinStorvikINLA} and~\citet{Friel2012}, which perform a comparison of a number of approaches to computing the marginal likelihood, including Laplace approximations, harmonic mean approximations, Chib's method, Chib and Jelizkov's method and INLA. The studies show that INLA is a fast method that enjoys giving precise approximations to the marginal likelihood even if the number of samples $T$ is limited.

\subsection{Mode jumping Markov chain Monte Carlo}

The main problem  with  the standard Metropolis-Hastings algorithms is the trade-off between  possibilities of large jumps (by which we understand proposals with a large neighbourhood) and  high acceptance  probabilities. Large jumps will typically result in  proposals with low probabilities.
In a continuous setting, \citet{Tjelmeland99modejumping} solved this  by  introducing  local optimization  after large jumps, which results in proposals with higher acceptance probabilities.  \citet{Hubin2016} adopted this approach to the discrete model selection setting and suggested the following algorithm:
\begin{algorithm}[H]
\caption{Mode jumping MCMC}\label{MJMCMCalg0}
\begin{algorithmic}[1]
\State  Generate a large jump  $\Bchi_0^*$ according to a  proposal distribution  $q_l(\Bchi_0^*|\Bgamma)$.
\item Perform a local  optimization, defined through $\Bchi_k^*\sim q_o(\Bchi_k^*|\Bchi_0^*)$.
\State Perform small randomization  to generate the proposal $\Bgamma^*\sim q_r(\Bgamma^*|\Bchi_k^*)$.
\item  Generate backwards auxiliary variables $\Bchi_0\sim q_l(\Bchi_0|\Bgamma^*)$,  $\Bchi_k\sim q_o(\Bchi_k|\Bchi_0)$.
\State Put
\[\Bgamma'=
\begin{cases}\Bgamma^*&\text{with probability $r_{mh}(\Bgamma,\Bgamma^*;\Bchi_k,\Bchi_k^*)$;}\\
\Bgamma&\text{otherwise,}
\end{cases}
\]
where
\begin{equation}
r_{mh}^*(\Bgamma,\Bgamma^*;\Bchi_k,\Bchi_k^*) = \min\left\{1,\frac{\pi(\Bgamma^*)q_r(\Bgamma|\Bchi_{k})}{\pi(\Bgamma)q_r(\Bgamma^*|\Bchi_k^*)}\right\}\label{locmcmcgen00}.
\end{equation}
\end{algorithmic}
\end{algorithm}
Here, a large jump corresponds to changing a large number of $\gamma_j$'s while the local optimization will be some iterative procedure based on, at each iteration, changing a small number of components until a local mode is reached. For this algorithm, three proposals need to be specified; $q_l(\cdot|\cdot)$ specifying the first large jump, $q_o(\cdot|\cdot)$ specifying the local optimizer, and $q_r(\cdot|\cdot)$ specifying the last randomization. All of them are described in detail in \citet{Hubin2016}. The convergence of the MJMCMC procedures is shown in Theorem~1 in \citet{Hubin2016}.  
\begin{figure}[ht]
\centering
\includegraphics[trim={0cm 0.98cm 1.0cm 1.8cm},clip,width=\linewidth]{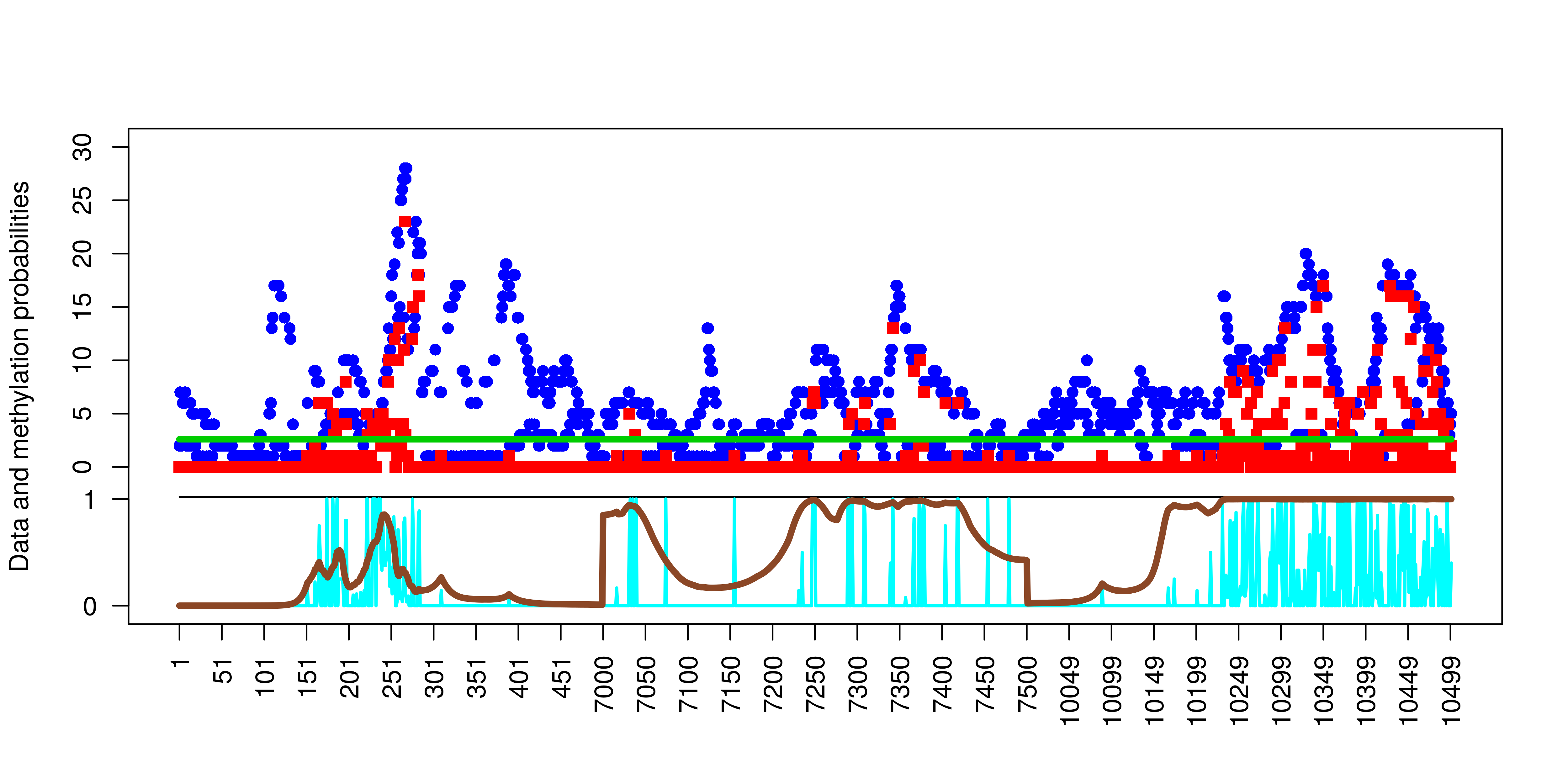}
\caption{\label{sample}Epigenetic observations, where blue dots are total number of reads, red dots - number of methylated reads, the green line corresponds to 2 total reads distinguishing the inference and the identification data, light blue line gives na\"ive probabilities as rates, brown line - probabilities as the posterior mean of the probability of success parameter from the posterior mode model. }
\end{figure}

\section{Data description}
The addressed data set consists of 1502 observations from chromosome one of the Arabidopsis plant belonging to five predefined groups of genes. This data set was divided into 950 observations (with more than 2 reads, see Figure \ref{sample}) for inference and 552 observations (with less than 3 reads) for model-based identification of methylation probabilities for the positions with the lack of data.

\begin{figure}[ht]
\centering
\includegraphics[trim={0.55cm 0.55cm 0.4cm 0.4cm},clip,width=0.68\linewidth]{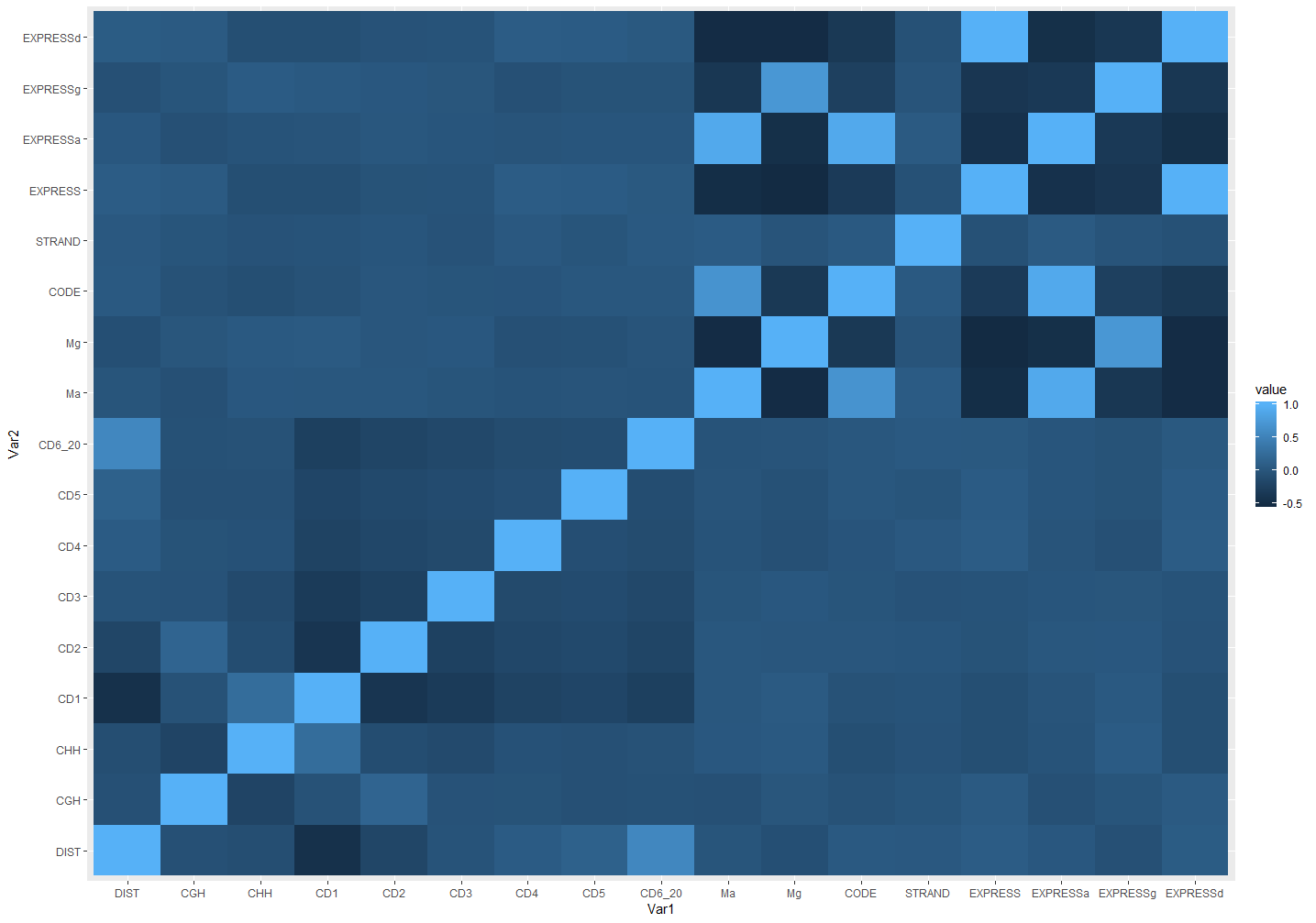}
\caption{Heat-map correlation plot between the addressed covariates.}\label{fig:corr}
\end{figure}

Apart from the observations represented by the methylated versus the total number of reads we have data on various exogenous variables (covariates). Among  these covariates, we address a factor with 3 levels corresponding to whether the location belongs to a CGH, CHH or CHG genetic region, where H is either A, C or T and thus generating two covariates $X_{CGH }$ and $X_{CHH}$. The second group of factors indicates whether the distance to the previous  cytosine nucleobase (C) in DNA is 1, 2, 3, 4, 5, from 6 to 20 or greater than 20 inducing six binary covariates $X_{DT1},X_{DT2},X_{DT3},X_{DT4},X_{DT5}$, and $X_{DT6:20}$. We also include such 1D distance as a continuous covariate $X_{DIST}$. The third addressed group of factors corresponds to whether the location belongs to a gene from a particular group of genes of biological interest. These groups are indicated as $M_a$, $M_g$ and $M_d$, yielding two additional covariates $X_{M_a},X_{M_g}$. Additionally, we have a covariate $X_{CODE}$ indicating if the corresponding nucleobase is in the coding region of a gene and a covariate $X_{STRD}$ indicating if the nucleobase is on a $"+"$ or a $"-"$ strand. Finally, we have a continuous covariate $X_{EXPR} \in \mathbb{R^{+}}$ representing expression level for the corresponding gene and interactions between expression levels and gene groups $X_{EXPR,a}, X_{EXPR,g}, X_{EXPR,d} \in \mathbb{R^{+}}$.  Thus multiple predictors with respect to a strict choice of the reference levels for categorical variables in our example induced $d = 17$ potentially important covariates. The correlation structure for the addressed variables is depicted in Figure~\ref{fig:corr}, where one clearly sees multiple correlations, which in turn are likely to induce multiple modes of the marginal likelihood.

\section{Results}
The MJMCMC algorithm was run until around $10\,000$ unique models ($7.6\%$ of the model space) were explored. We parallelized the search on 10 CPUs. Default tuning parameterrs from \cite{Hubin2016} were used.

% The following R call is used for running the inference:

% \begin{lstlisting}[language=R]
% fib <- function(n) {
%   if (n < 2)
%     n
%   else
%     fib(n - 1) + fib(n - 2)
% }
% fib(10) # => 55
% \end{lstlisting}

\begin{figure}[ht]
\centering
\includegraphics[trim={0cm 1.65cm 1.0cm 1.8cm},clip,width=0.68\linewidth]{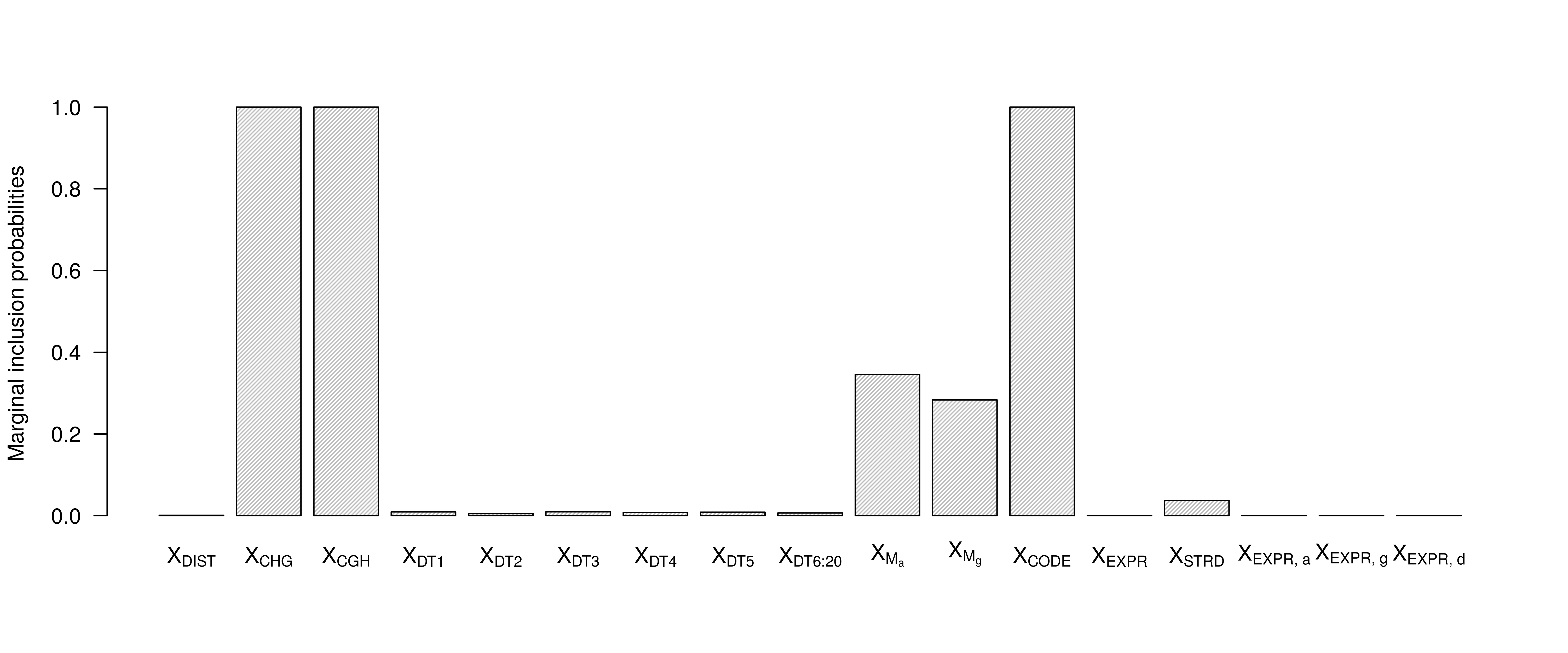}
\caption{Barplots of RM estimates \citep{Hubin2016} of marginal inclusion probabilities of the covariates.}\label{margin}
\end{figure}

According to the marginal inclusion probabilities reported in Figure~\ref{margin}, only three factors $X_{CHG}, X_{CGH}$ and $X_{CODE}$ are clearly significant for inference on the methylation patterns for the addressed epigenetic region, factors $X_{M_a}$ and $X_{M_g}$ also have some significance. 
Table~\ref{tablebest} gives the marginal posterior model probability and posterior means of the parameters for the best model in the explored subset of models from the model space. This model is both the posterior mode model in the set of explored models and the median probability model \citep{barbieri2004optimal}. We have also compared the selected model with alternative models based on the optimal sets of covariates but with other latent Gaussian structures and found our model to be the best in terms of the marginal log likelihood (see the Appendix A of the paper). %The results summarized in Table~\ref{latproc} show that also for the selected model our choice of the structure of the latent Gaussian field is optimal (on the set of explored alternatives).

\begin{table}[ht]
\centering
\begin{tabular}{ 
c|cccccc}
\hline
PMP&$\beta_0$&$\beta_{CHG}$&$\beta_{CGH}$&$\beta_{CODE}$&$\tau_{\epsilon}$&$\tau_{\zeta}$\\\hline
0.4276&-8.8255&2.4717&5.2122&6.4240&7.5075&1.2109\\ 
\hline
\end{tabular}\
\\[1pt]
\caption{Posterior means for the best model in terms of marginal posterior probability (PMP).}\label{tablebest}
\end{table}
% \begin{figure}[ht]
% \centering
% \includegraphics[trim={0cm 0.98cm 1.0cm 1.8cm},clip,width=0.68\linewidth]{latent.png}
% \caption{\label{sample2}Posterior means of the latent process for the locations in the studied genetic region. }
% \end{figure}
Based on the best model, we carried out computations of methylation probabilities for the locations in both the inference set and the identification set. Highly methylated regions are located between observations 7000-7050, 7250-7400, and 10150-10500, see Figure~\ref{sample}. Note that the model is quite sceptical to the methylation status of locations 7051 - 7249, despite a number of observations with a high proportion of methylated reads in this region. Furthermore, we compared the results with the na\"ive approach based on computing the proportion of methylated reads (light blue line in Figure~\ref{sample}), which is currently addressed in the biological literature as a standard way to evaluate methylation probability of a given nucleobase. The results show that the na\"ive approach should not be trusted in the presence of spatially correlated data and the probabilities corresponding to it  can be strongly biased. 

\section{Discussion}

During cancer development, the changes in DNA methylation patterns occur within the gene promoter, CpG islands and their shores \citep{tang2017integration}. Hence in future, it would be of interest to obtain additional covariates such as whether the corresponding nucleotide base position belongs to a particular part of the non-coding gene region like a promoter, an intron or remnants of transposable elements, and whether the nucleobase is within a CpG island, and see how these covariates are influencing the underlying methylation patterns. At the same time, in this paper we looked only at a subset of the genomic locations associated with the groups of genes of biological interest, however, in the future, it would be of interest to address the whole genome. That would induce working with extremely large data, which in turn creates new methodological challenges. In particular, in order to make the efficient inference, it will be important to allow  sub-sampling for INLA within a given model and MJMCMC in the discrete marginal space of models. It will also be of interest to allow logical expressions for the binary covariates to be included in the model following \citet{hubin2018novel}. Finally, joint inference on the covariates and various latent Gaussian variables by means of MJMCMC can be of interest.

\small{
\subsection*{Acknowledgements}
The authors would like to thank the Center for Evolutionary Life Sciences (CELS) at the University of Oslo and BBPUBL19 project at the Norwegian Computing Center for fully financially supporting this work. Additionally, we thank Dr Olav Nikolai Breivik (Norwegian Computing Center) for thoroughly proofreading the article.
}
\bibliography{guidelinesAJS}

\appendix
\section{Appendix}
\subsection{Alternative latent models}
We also looked at several alternative latent Gaussian variables. These are reported in Table~\ref{latproc} together with the marginal likelihoods (Bayes factors can be computed in a straight-forward fashion). The default priors for the parameters from INLA~\citep{rue2009eINLA} were used. Here, we clearly see that the chosen structure of the latent Gaussian field is significantly outperforming all other alternatives for both the null and the full models, which gives strong evidence to our choice. This is also true for the model selected by MJMCMC, which is reported in Table~\ref{tablebest}.
\begin{table}[h]
\begin{center}
\begin{tabular}{lrrrrr}
\hline
${IG}$ +&${RW(1)}$&${OU}$&${AR(1)}$&${AR(2)}$&${AR(3)}$\\\hline%&AR(4)&AR(5)&AR(6)&AR(7)&AR(8)\\\hline
FULL&\textbf{-771.7173}&-1103.8190&-1116.8367&-1103.8188   &-811.2963\\\hline%&-876.8078 &-1114.0377&-1280.6415 &-815.6523  &-1121.9294\\\hline
NULL&\textbf{-875.4917}&-876.5565&-876.8325&-876.5565&-875.4917\\\hline%&-na&-873.1148&-875.1247\\\hline%&-na&-na\\\hline 
BEST&\textbf{{-591.0837}}&-706.8185&-713.7015&-706.8185&-697.3347\\\hline%&-na& -na&-1025.5220&-697.7801&-695.6773 \\\hline
\end{tabular}
\end{center}
\caption{\small Estimates of the marginal log likelihood for different candidate latent processes, which are Random Walk ($RW(1)$), Ornstein–Uhlenbeck (${OU}$) and auto-regressive processes $AR(p)$ of order $p$ from 1 to 3. Here \textit{FULL} is the model with all covariates included, \textit{NULL} is the model without covariates, and \textit{BEST} is the model with the covariates from Table~\ref{tablebest}.}\label{latproc}
\end{table}

\end{document}